\documentclass[journal]{IEEEtran}
\ifCLASSINFOpdf
   \usepackage[pdftex]{graphicx}
  \graphicspath{{./drawing1/}}
   \DeclareGraphicsExtensions{.pdf,.jpeg,.png}
\else
\fi
\begin{document}
%
\title{Development and Characterization of a 3.2 Gb/s Serial Link Transmitter for CMOS Image Sensors in Subatomic Physics Experiments}
%
%
%

\author{Quan~Sun,
        Guangyu~Zhang,
        Datao~Gong,
        Binwei~Deng,
        Wei~Zhou,
        Bihui~You,
        Le~Xiao,
        Jian~Wang,
        Dongxu~Yang,
        Tiankuan~Liu,
        Chonghan~Liu,
        Di~Guo,
        Jun~Liu,
        Christine~Hu-Guo,
        Frederic~Morel,
        Isabelle~Valin,
        Xiangming~Sun,
        and~Jingbo~Ye
\thanks{Q. Sun is with the Department of Physics, Southern Methodist University, Dallas, TX, 75205 USA(e-mail: quans@smu.edu).}
\thanks{G. Zhang, J. Wang and D. Yang are with University of Science and Technology of China, Hefei Anhui 230026, P.R. China.}
\thanks{D. Gong, T. Liu, C. Liu and J. Ye are with the Department of Physics, Southern Methodist University, Dallas, TX, 75205 USA.}
\thanks{B. Deng is with Hubei Polytechnic University, Huangshi, Hubei 435003, P.R. China.}
\thanks{W. Zhou, B. You, L. Xiao, D. Guo, J. Liu and X. Sun are with Department of Physics, Central China Normal University, Wuhan, Hubei 430079, P.R. China.}
\thanks{C. Hu-Guo, F. Morel, I. Valin are with Institut PluridisciplinaireHubert Curien, CNRS/IN2P3/UDS 23 rue du loess, BP 28, 67037 Strasbourg, France}
\thanks{Manuscript received .}}

%
%

\markboth{ }%
{Shell \MakeLowercase{\textit{et al.}}: Bare Demo of IEEEtran.cls for IEEE Journals}
%



\maketitle

\begin{abstract}
This paper presents development and characterization of a 3.2 Gb/s serial link transmitter for CMOS image sensors. The transmitter incorporates Reed-Solomon code to achieve low error rate in the harsh environment of subatomic physics experiments. Pre-emphasis is implemented in the transmitter, allowing data transmission over low-mass cables. It is fabricated in a 0.18 $\mu m$ CMOS image process as a standalone chip to characterize its performance, with the core area of 1.8 $mm^2$. A frame data rate of $3\cdot10^{-12}$ with confidence level of 94.5$\%$ was measured through a FPGA based receiver. The measured nominal power consumption is 135 mW. The transmitter functions normally after irradiated with 4.5 Mrad TID.
\end{abstract}


\hyphenation{op-tical net-works semi-conduc-tor}

%
\IEEEpeerreviewmaketitle

\section{Introduction}
%
%
%
%
\IEEEPARstart{C}{MOS} monolithic active pixel Sensors (MAPS) have demonstrated good performance for tracking devices\cite{1748-0221-4-04-P04012,Hu-Guo-1234604,BAUDOT2013480} and are demanded by numerous future subatomic physics experiments. They integrate sensing elements, signal processing and readout electronics on a single chip, providing a trade-off among granularity, material budget, power consumption and readout speed.

Most of current MAPS employ parallel data links for data transmission, which provide an efficient and robust data transmission at a low data rate. Despite its simple structure, parallel data transmission suffer from two major issues, clock skew, and interference. As increasing of hit density of subatomic physics experiments, both of the issues are becoming notable. Material budget is another considerable limitation for a parallel data link. Cables used in parallel data transmission contribute material significantly. By contrast, serial data link allows high-speed data transmission with only a pair of differential cable, considerably reduce material in the detector system. Therefore, serial data transmission show strong potential for future MAPS application in subatomic physics experiments.

Although widely used in industry applications, few serial link transmitters are dedicated to MAPS application in subatomic physics experiments. An early try is a 160 Mb/s one which employs 8b10b encoding to provide DC balanced data\cite{qsun2009}. Recently, a 1.2-Gb/s serial data transmitter dedicated to MAPS data transmission is developed and characterized for ALICE detector upgrade\cite{cern-twepp2016}, which employs triple modular redundancy (TMR) technique in the clock tree and the serializer to protect from single event transient (SET). It also use 8b10b code to provide DC balance and error detection abilities. 

In this paper, we present the development and characterization of a 3.2-Gb/s serial link transmitter for MAPS data transmission in future subatomic physics experiments. The transmitter features employment of Reed-Solomon coding technique, therefore allowing some errors in the transmission to be recovered at the receiving end. Double data rate (DDR) scheme is adopted by the transmitter to operate at a low clock frequency. In order to send data through low-mass cables, pre-emphasis is incorporated in the current mode logic (CML) driver. The transmitter is implemented in a 0.18 $\mu$m CMOS technology.

In the next section, the transmitter architecture is detailed. Section 3 presents the development of building blocks. Section 4 describes the electrical and radiation test results of the prototype. Finally, conclusions are drawn in section 5.

 

\section{the transmitter architecture}
\begin{figure}[!t]
	\centering
	\includegraphics[width=1\linewidth]{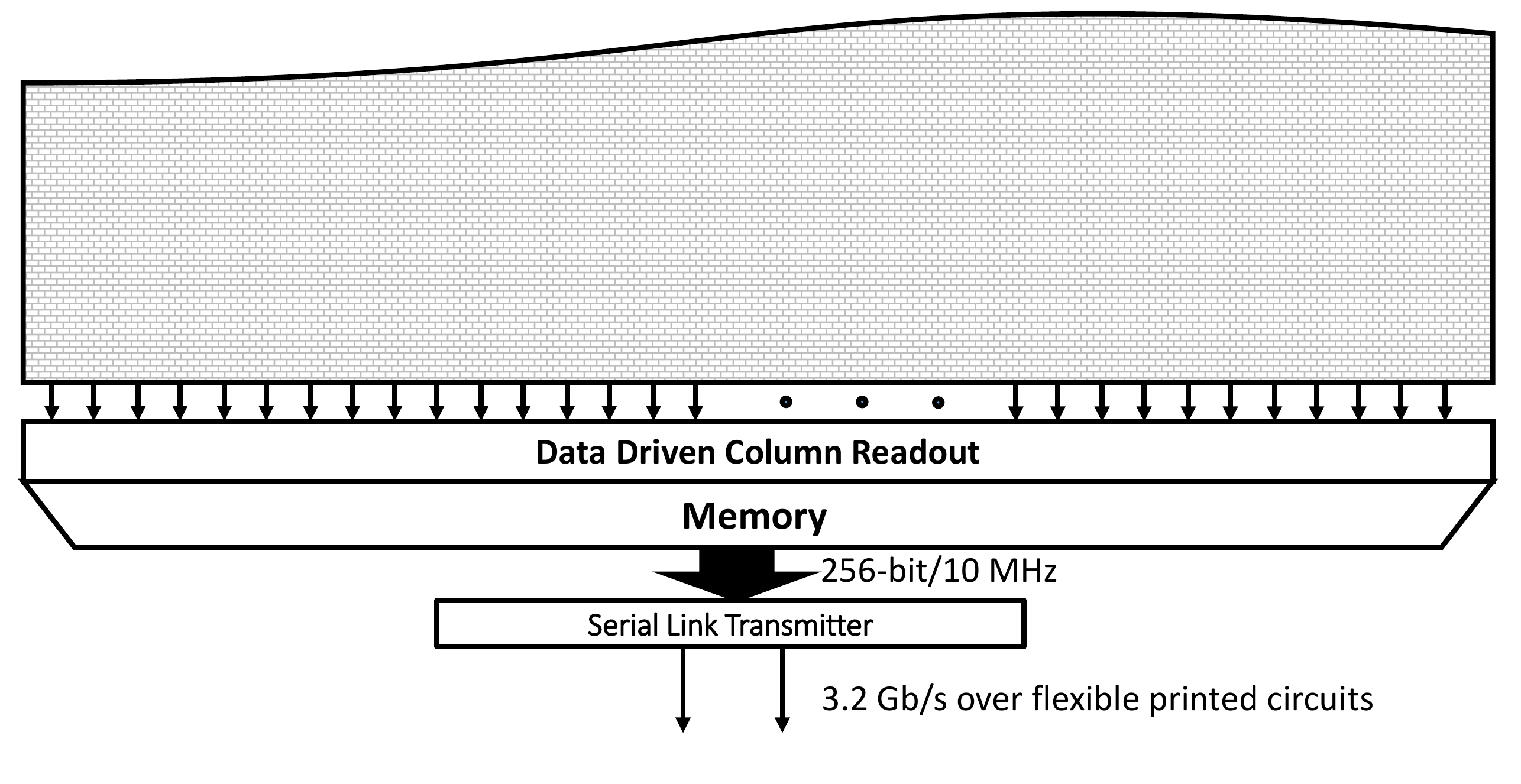}
	\caption{The architecture of MAPS integrated the transmitter.}
	\DeclareGraphicsExtensions{.pdf,.jpeg,.png}
	\label{maps}
\end{figure}

The block diagram of a typical MAPS chip foreseen to equip the transmitter is shown in Fig. \ref{maps}. Particle hits in the pixel array are read out to the bottom of the chip by data-driven readout circuits in column level, and then some kind of zero suppression technology might be employed to compress raw data. A succeeding chip level memory is used to provide several frames storage capability to buffer the frame data. The final frame data are formatted to 256-bit width at 10 MHz, which are sent out by the proposed serial link transmitter finally through a 100 $\Omega$ differential transmission line on a flexible printed circuits with the length of up to several tens centimeter. 

\begin{figure}[!t]
	\centering
	\includegraphics[width=1\linewidth]{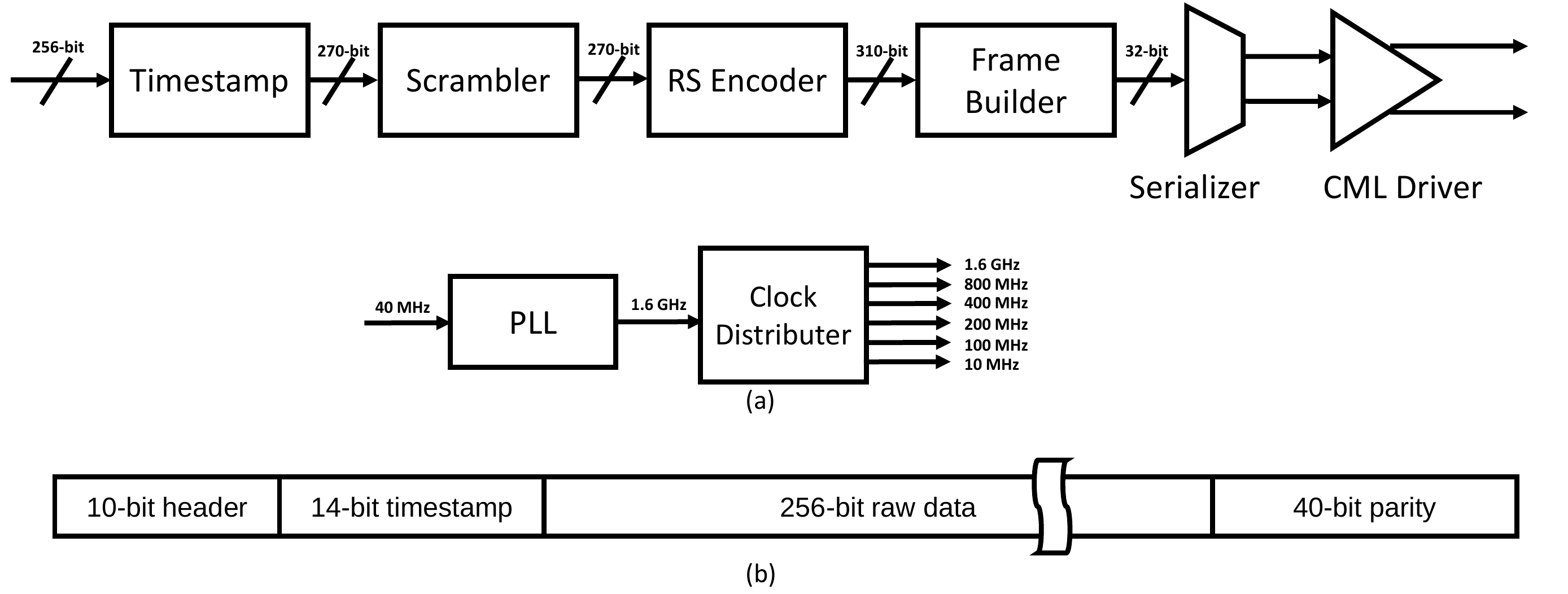}
	\caption{(a)the transmitter architecture, (b)frame definition.}
	\DeclareGraphicsExtensions{.pdf,.png}
	\label{arch}
\end{figure}

Fig. \ref{arch}(a) illustrates a simplified block diagram of the proposed transmitter. The 256-bit raw frame data are firstly added a 14-bit timestamp. The scrambler transposes the 270-bit full frame data including time information to provide DC balance, which enables receivers to recover clock from the serial data. The scrambled data are encoded with a Reed-Solomon algorithm which adds 40-bit redundancy into the data so that errors can be corrected with a decoder in receivers. In the succeeding frame builder, a 10-bit frame header is added in each frame, which allows synchronizing the frame in receivers. The frame building also act as a low-speed serializer which multiplexes each 320-bit frame with a 100 MHz clock, generating 32-bit data for the succeeding high-speed 32:1 serializer. The serializer multiplexes the 32-bit data and provides a pair of complementary signal and its two copies of one and two bit-period delay to the succeeding CML driver. The CML driver which incorporates pre-emphasis delivers serial data at 3.2 Gb/s finally. Several clocks are required in the above-mentioned operation, i.e. 10 MHz, 100 MHz, 200 MHz, 400 MHz, 800 MHz and 1.6 GHz. A PLL generates the 1.6 GHz clock from a 40 MHz reference clock, and then all the required clocks are generated from the 1.6 GHz PLL clock by a clock distributer. In Fig. \ref{arch}(a), all the blocks preceding the serializer are implemented with a digital IC flow and triplicated to resist SEU(Single Event Upset). Two TMR (Triple Module Redundancy) schemes are applied, say, all the registers are triplicated locally, while all the combinational logic are triplicated and voted globally at the output of the frame builder. The frame structure is shown in Fig. \ref{arch}(b), in which a frame includes a 270-bit payload and a 50-bit overhead.

\section{building blocks}
\subsection{The Scrambler}
\begin{figure}[!t]
	\centering
	\includegraphics[width=1\linewidth]{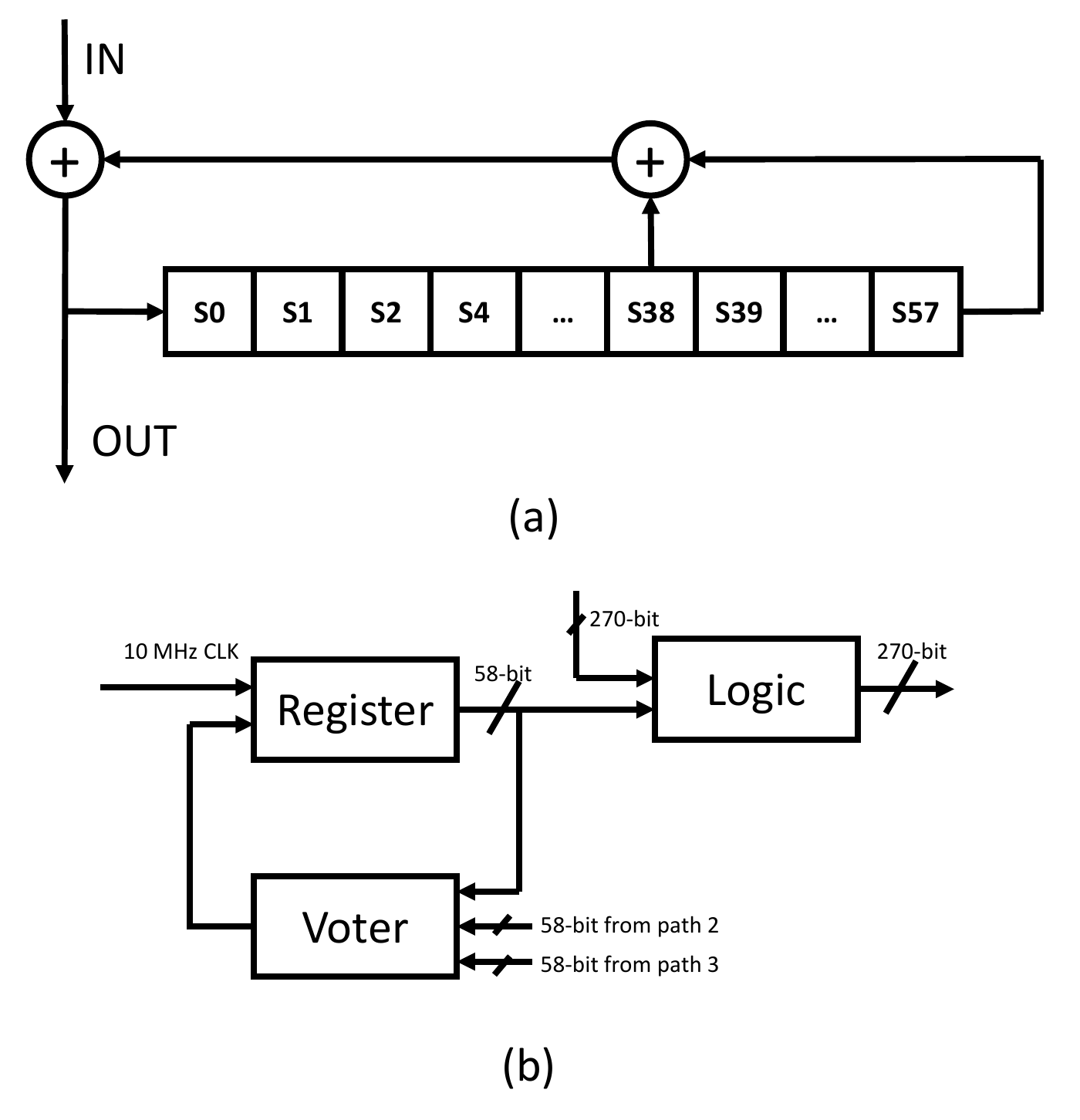}
	\caption{Scrambler implementing the polynomial of $x^{58}+x^{39}+1$: (a)a conventional serial implementation of scrambler, (b)parallelized implementation of the scrambler with local voter.}
	\DeclareGraphicsExtensions{.pdf,.jpeg,.png}
	\label{scr}
\end{figure}

The scrambler is employed to provide a DC balanced data, which ease clock recovery in receivers. The scrambling polynomial $x^{58}+x^{39}+1$ is implemented in the scrambler. Fig. \ref{scr}(a) shows a conventional scrambler implementation which scrambles data serially at bit-clock. This structure directly maps to the polynomial and therefore is easy to understand.However, for the scenario in the proposed transmitter, a 2.7 GHz bit clock is required if the structure in Fig. \ref{scr}(a) is used due to the 270-bit input data at 10 MHz clock. To avoid the using of 2.7 GHz clock, a parallelized implementation is employed to generate 270-bit scrambled data at 10 MHz frame clock rate. Fig. \ref{scr}(b) illustrates the parallelized implementation of the scrambler. A local TMR loop is implemented to protect register from SEU. A combinational logic block generates 270-bit scrambled data from the input and the register. The protection for the combinational logic circuits is considered globally.

\subsection{The RS Encoder}
Reed-Solomon codes are block-based error correcting codes with a wide range of applications in digital communications and storage. A Reed-Solomon code is specified as RS(n,k) with s-bit symbols, meaning that the encoder takes k data symbols of s bits each and adds parity symbols to make an n symbols codeword. A Reed-Solomon decoder can correct up to t symbols that contain errors in a code word, where $t=(n-k)/2$.

\begin{figure}[!t]
	\centering
	\includegraphics[width=1\linewidth]{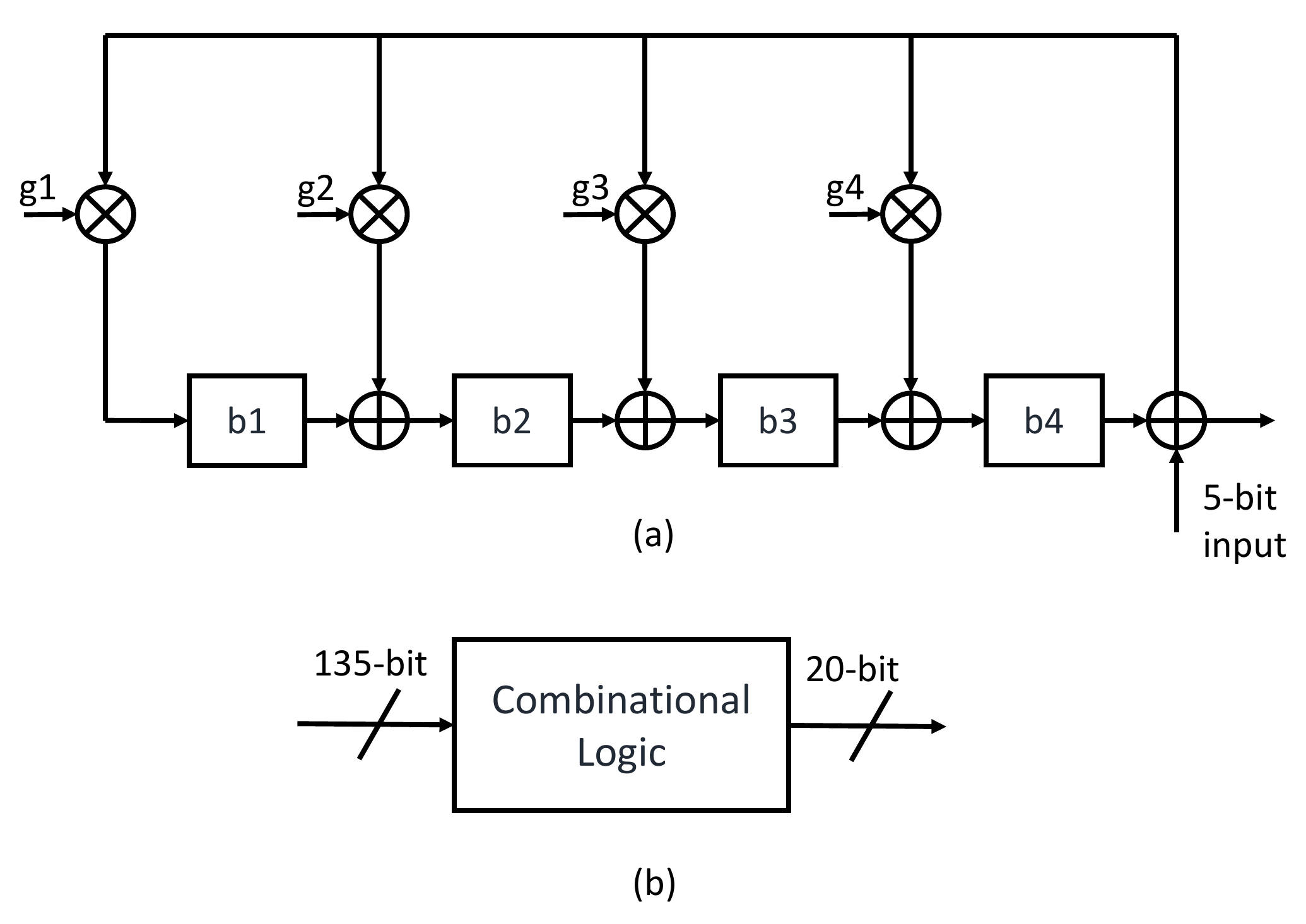}
	\caption{Reed-Solomon encoder:(a)conventional implementation with Galois field arithmetic, (b)parallelized implementation in the transmitter}
	\label{rs}
\end{figure}

In the proposed transmitter, a Reed-Solomon code of RS (31, 27) with 5-bit symbols is chosen to trade off between the ability of correction and overhead of transmission. An interleaved encoding scheme is adopted to optimize burst error correction and double the encoding rate. Two identical encoders are interleaved in the transmitter, and each of the encoder processes 135-bit data. The implementation of the encoder is illustrated in Fig. \ref{rs}. Fig. \ref{rs}(a) is a conventional encoder which generates parity bits under the control of the clock at the symbol rate, 270 MHz in our case. Galois field arithmetic is used in the encoder. Coefficients of polynomial, g1, g2, g3, and g4, are calculated from generator polynomial equation \ref{eq_rs}.

\begin{equation}\label{eq_rs}
g(x)=(x-\alpha^{27})(x-\alpha^{28})(x-\alpha^{29})(x-\alpha^{30}).
\end{equation}

In the structure of Fig. \ref{rs}(a), since additional timing is required to handle the payload, a clock of 320 MHz has to be applied. In our case, we parallelized the Galois field arithmetic by using combinational logic, which allows the encoder operating at 10 MHz frame clock, as shown in Fig. \ref{rs}(b)\cite{gf-rs}. The encoders are triplicated globally.

\subsection{The Serializer}
\begin{figure}[!t]
	\centering
	\includegraphics[width=1\linewidth]{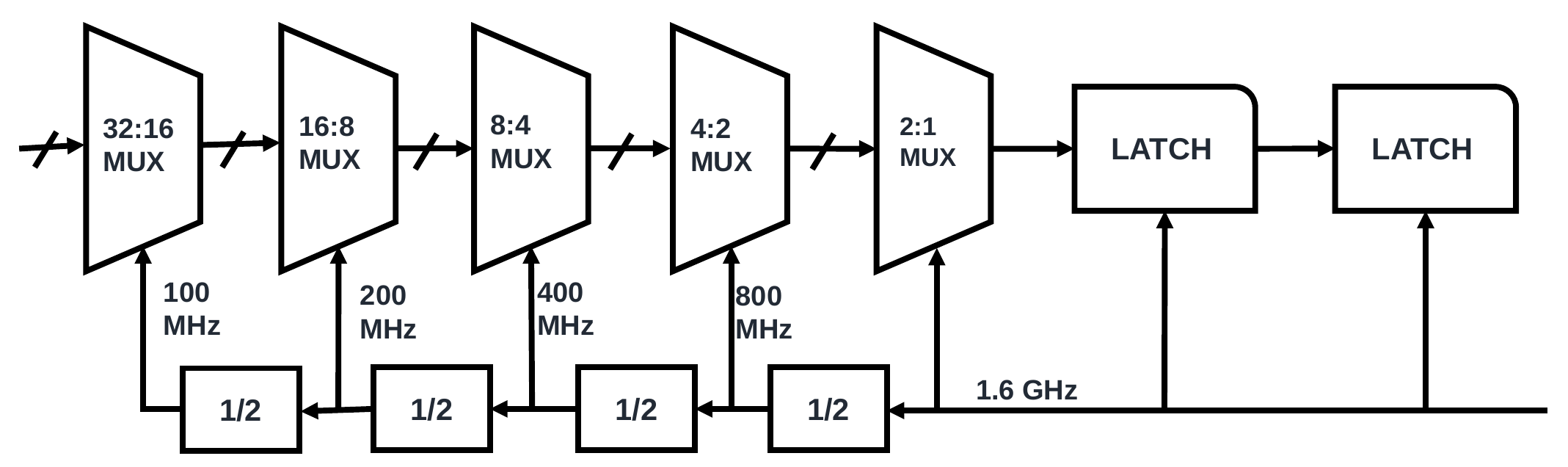}
	\caption{stucture of the serializer in the transmitter}
	\DeclareGraphicsExtensions{.pdf,.jpeg,.png}
	\label{serializer}
\end{figure}

The serializer includes a full-custom 32:1 multiplexer with 5-stage binary-tree structure and two additional latches, as shown in Fig. \ref{serializer}. Five clocks are provided by the PLL and the clock distributer. The first stage multiplexer interfaces with frame builder at 100 MHz clock. The last stage of the multiplexer and two latches, working at 1.6 GHz clock, provides a complementary signal and its two copies of one and two bit-period delay to the CML driver. The 50\% duty cycle of the 1.6 GHz is critical because of the DDR operation.
Since the data to be serialized have been protected by the Reed-Solomon codes, no TMR is applied in the serializer, which ease the design of the serializer and minimize its power consumption.

\subsection{The CML Driver with Pre-emphasis}
\begin{figure}[!t]
	\centering
	\includegraphics[width=1\linewidth]{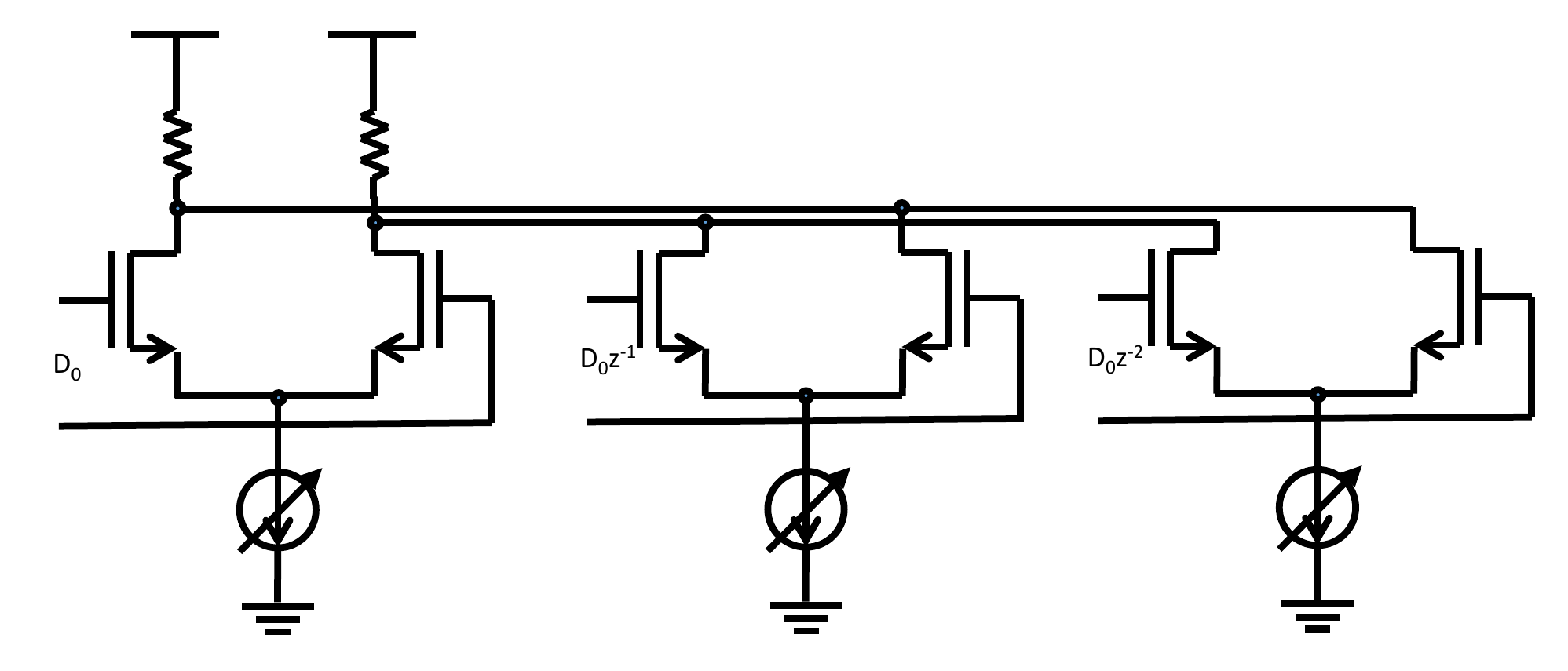}
	\caption{simplified schematic of the CML driver with pre-emphasis}
	\DeclareGraphicsExtensions{.pdf,.jpeg,.png}
	\label{cml}
\end{figure}

In order to meet tight material budget requirement of subatomic physics experiments, low-mass traces have to be used for data transmission. The proposed transmitter employs pre-emphasis in the driver, which provides compensation to the loss of the traces. The CML driver in the transmitter incorporates two post taps to realize pre-emphasis, as shown in Fig. \ref{cml}. The additional two taps are driven by the delayed copies of the dominant tap. A Z-domain transfer function implemented by the driver is shown in equation \ref{eq_cml}. Two parameters, $a_0$ and $a_1$, are programmable through a slow-control circuits.
\begin{equation}\label{eq_cml}
Y(z^{-1})=I_o(1+a_0z^{-1}+a_1z^{-2}).
\end{equation}

\subsection{The Clock Genration Circuits}
\begin{figure}[!t]
	\centering
	\includegraphics[width=1\linewidth]{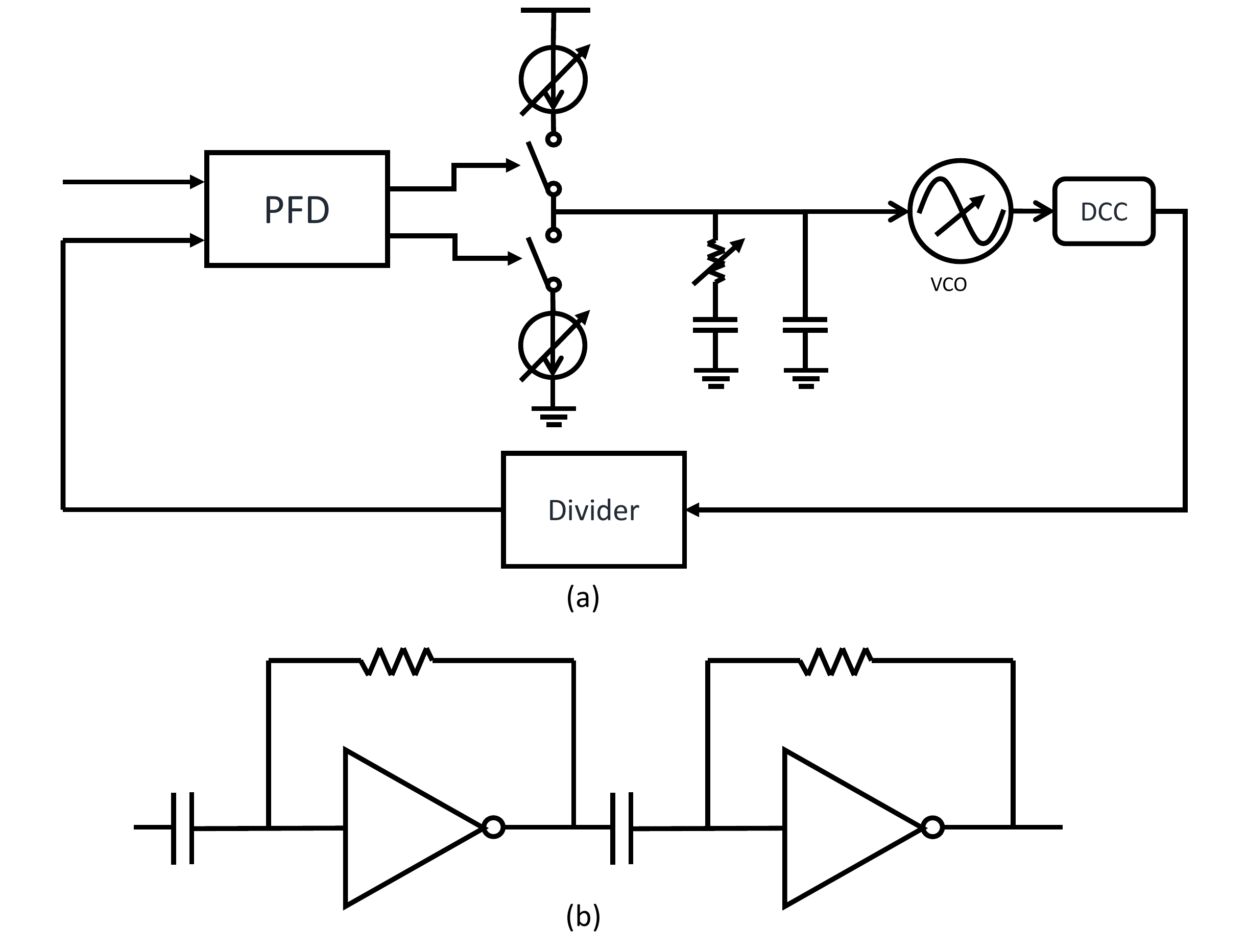}
	\caption{the PLL block diagram used in the transmitter}
	\DeclareGraphicsExtensions{.pdf,.jpeg,.png}
	\label{pll}
\end{figure}

A charge-pump PLL is used to generate the 1.6 GHz clock from a 40 MHz reference clock, as shown in Fig. \ref{pll}(a). The loop bandwidth is programmable from 0.5 MHz to 2 MHz, which eases the bandwidth trade-off between in-band noise and out-band noise in field applications. A 4-stage ring VCO(Voltage-Controlled Oscillator) is designed to cover frequency ranging from 0.8 GHz to 2.4 GHz. The divider in the PLL is fully triplicated to tolerate SEU.
A duty cycle correction block, shown in Fig. \ref{pll}(b), is applied to minimize DCD (duty cycle distortion) by providing a duty cycle of near 50\%. This circuit correct duty cycle by storing offset of the clock and the inverters on the capacitors.


\section{Characterization of the Transmitter}
\begin{figure}[!t]
	\centering
	\includegraphics[width=1\linewidth]{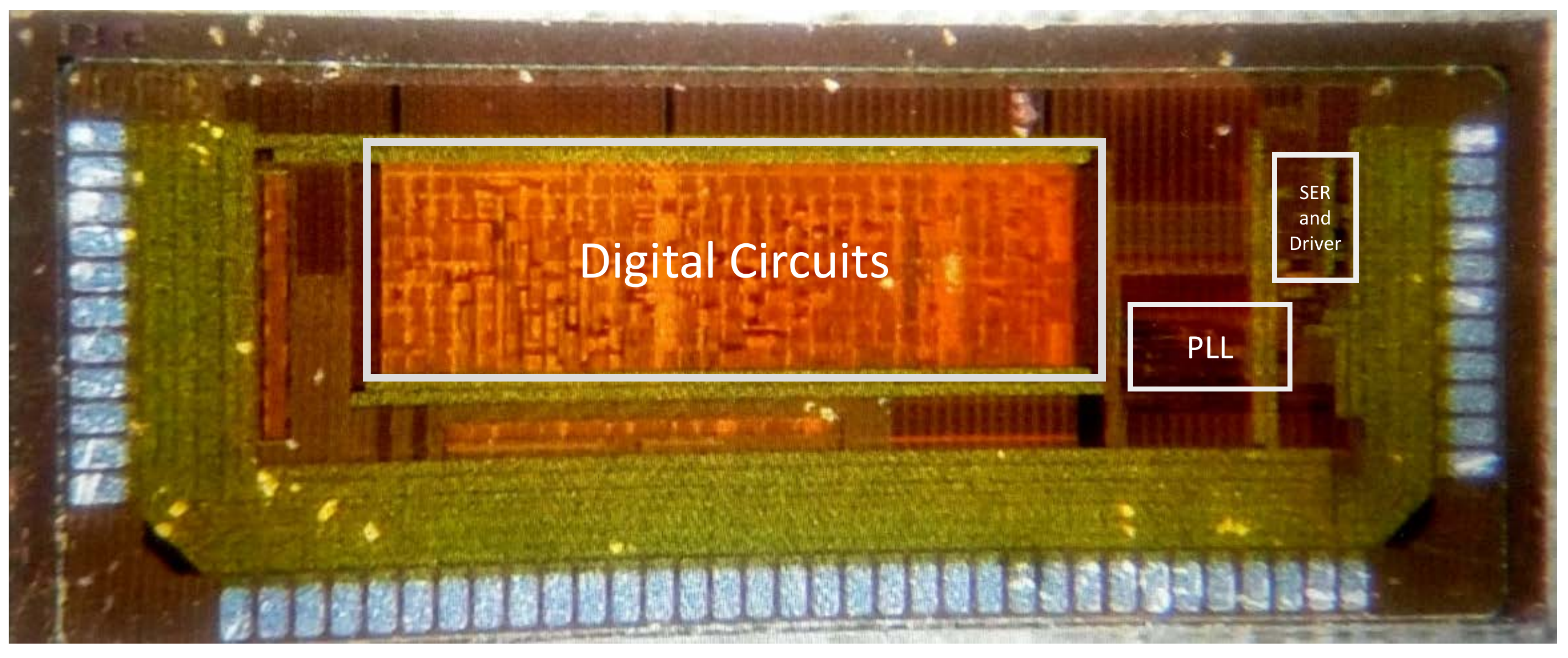}
	\caption{photomicrograph of the transmitter.}
	\DeclareGraphicsExtensions{.pdf,.jpeg,.png}
	\label{die}
\end{figure}

The transmitter is fabricated in a 0.18 $\mu$m CMOS image sensor technology. The core of the prototype is 2360 $\mu$m by 760 $\mu$m. Fig. \ref{die} shows the photomicrograph of a die and block description. The die was wire-bonded on a four-layer FR4 PCB(Printed Circuits Board) for characterization in laboratory.

\subsection{Electrical Test Results}
\begin{figure}[!t]
	\centering
	\includegraphics[width=1\linewidth]{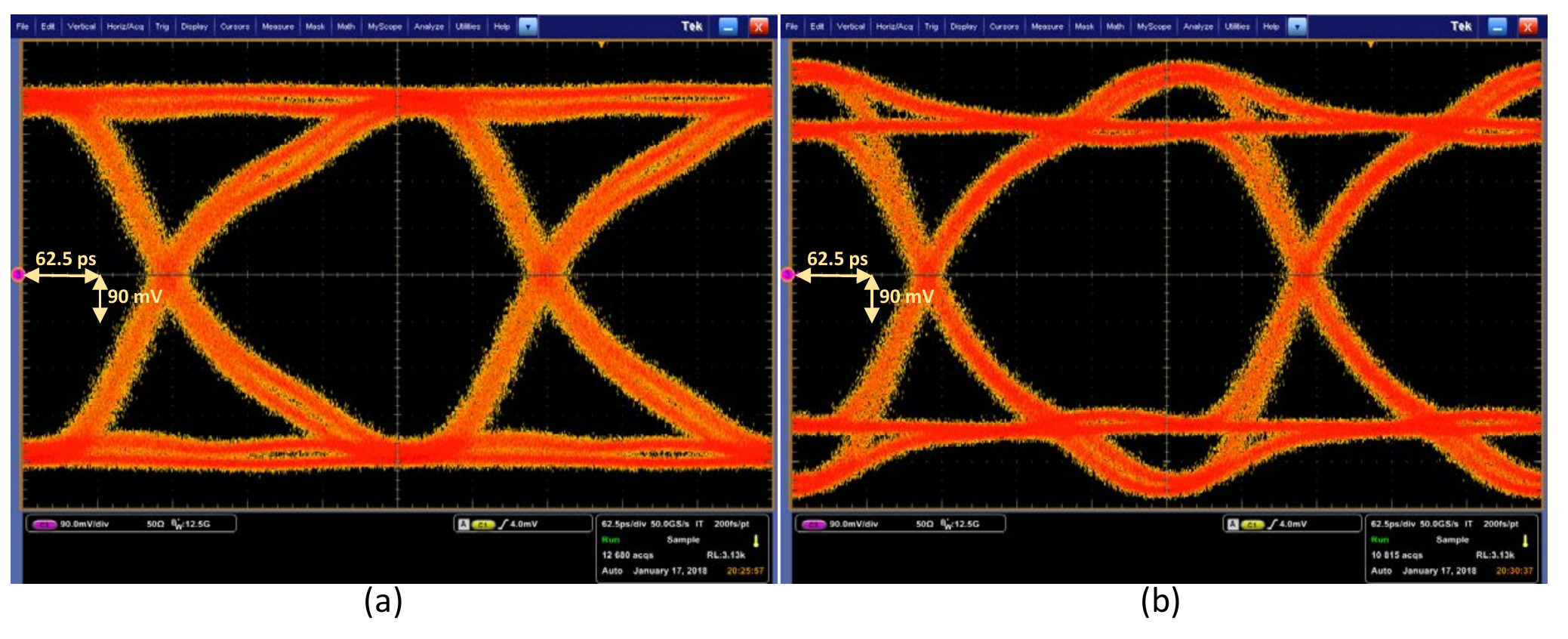}
	\caption{eye diagrams measured after two RF25S cables with 1 m length, (a) pre-emphasis off, (b) pre-emphasis on with $a_0=-0.2$ and $a_1=0$.}
	\DeclareGraphicsExtensions{.pdf,.jpeg,.png}
	\label{eye}
\end{figure}

To measure the quality of the serial data sent by the transmitter, a 40 MHz reference clock is provided by a clock generator, SI5338 clock generator evaluation board. A Tektronix DSA71254B digital serial analyzer is used to measure the eye diagram and jitter performance of the serial data. The measured jitters are summarized in Table \ref{tab_jitter}. Fig. \ref{eye} shows eye diagrams, with pre-emphasis off and on. 

A serial data receiver matching the proposed transmitter has been developed on a FPGA evaluation board(KC705) and used in the measurement. Benefitting from Reed-Solomon coding, most of data errors arising after the encoder could be corrected or detected by the receiver. Three outcomes are provided by the receiver:
\begin{itemize}
	\item Count of incorrect frames before Reed-Solomon decoding, which represent transmission errors usually.
	\item Count of payloads failed to be corrected but are detected by the receiver.
	\item Count of incorrect payloads after decoding and de-scrambling, which is generated by comparing with original payloads to be sent.
\end{itemize}

It is possible that the receiver fails to detect an incorrect frame or incorrectly decode an payload without any warning. The third error indication in above list will show this situation. The probability of each above case depends on the number and distribution of errors. 

To measure error rate, the outputs of the transmitter were directly connected to the receiver with low-loss cable to perform measurements. For this direct electrical connection, no error in any of above case is detected for a 27 hours measurement, corresponding to a FER(Frame Error Rate) of $3\cdot10^{-12}$ with confidence level of 94.5$\%$, or to a BER (Bit Error Rate) for the payload of $1.2\cdot10^{-14}$ with confidence level of 94.9$\%$.

\begin{figure}[!t]
	\centering
	\includegraphics[width=1\linewidth]{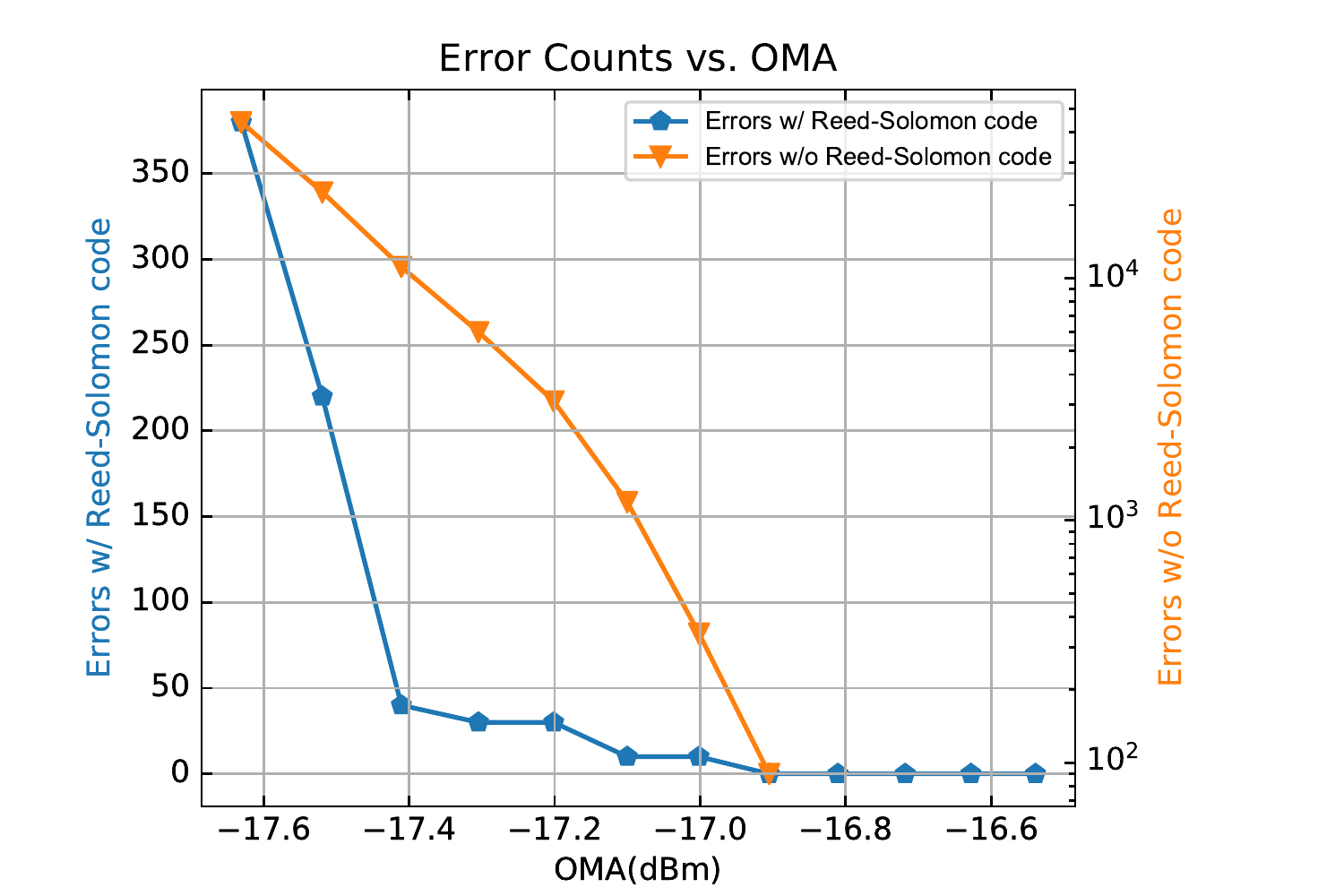}
	\caption{measured error counts with transmission errors added. Frame Error Counts means number of incorrect payload after correction, while Transmission Error Counts indicates number of incorrect frames detected before correction. }
	\DeclareGraphicsExtensions{.pdf,.jpeg,.png}
	\label{errorcount}
\end{figure}

With the help of the above three indications from the receiver, one could observe the frame errors with and without the Reed-Solomon code. In order to investigate the benefit of the Reed-Solomon code, some errors are intentionally added during transmission by inserting an optical transceiver, FINISAR FTLX8574D3BCL, and a variable optical attenuator into the data link. Controlled transmission errors are generated by attenuating optical power to a range around the optical sensitivity of the optical receiver. Fig. \ref{errorcount} shows measured results with transmission errors. Each measurement time is 10 minutes, corresponding to a frame counts of $6\cdot10^9$. The transmission excluding the Reed-Solomon coding is error-free when the optical modulation amplitude(OMA) is above -16.8 dBm, while errors increase rapidly with decreasing of the OMA. The recovered payloads are error-free when the OMA is above -16.9 dBm. At the OMA of -16.9 dBm, 90 incorrect frames are correctly recovered. Detailed analysis shows that most of the frame errors after correction come from the failures of frame synchronization. That means the receiver cannot find headers due to errors in them. A single bit error in the header will generate 10 consecutive frame errors after the decoder. This problem could be avoided by optimizing header detection of the receiver.

\subsection{Irradiation Test Results}
\begin{figure}[!t]
	\centering
	\includegraphics[width=1\linewidth]{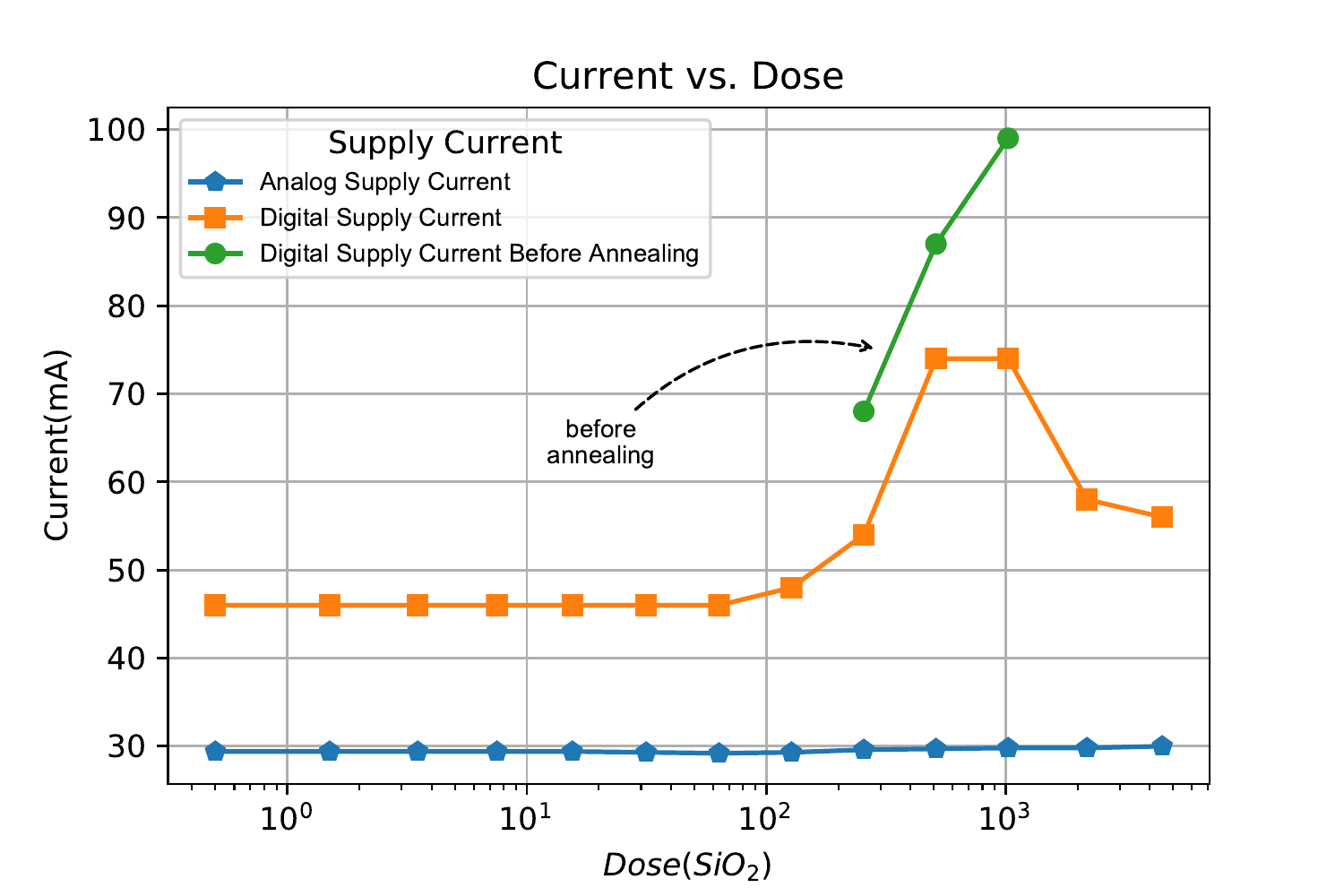}
	\caption{current consumption vs. TID.}
	\DeclareGraphicsExtensions{.pdf,.jpeg,.png}
	\label{radcurrent}
\end{figure}

\begin{figure}[!t]
	\centering
	\includegraphics[width=1\linewidth]{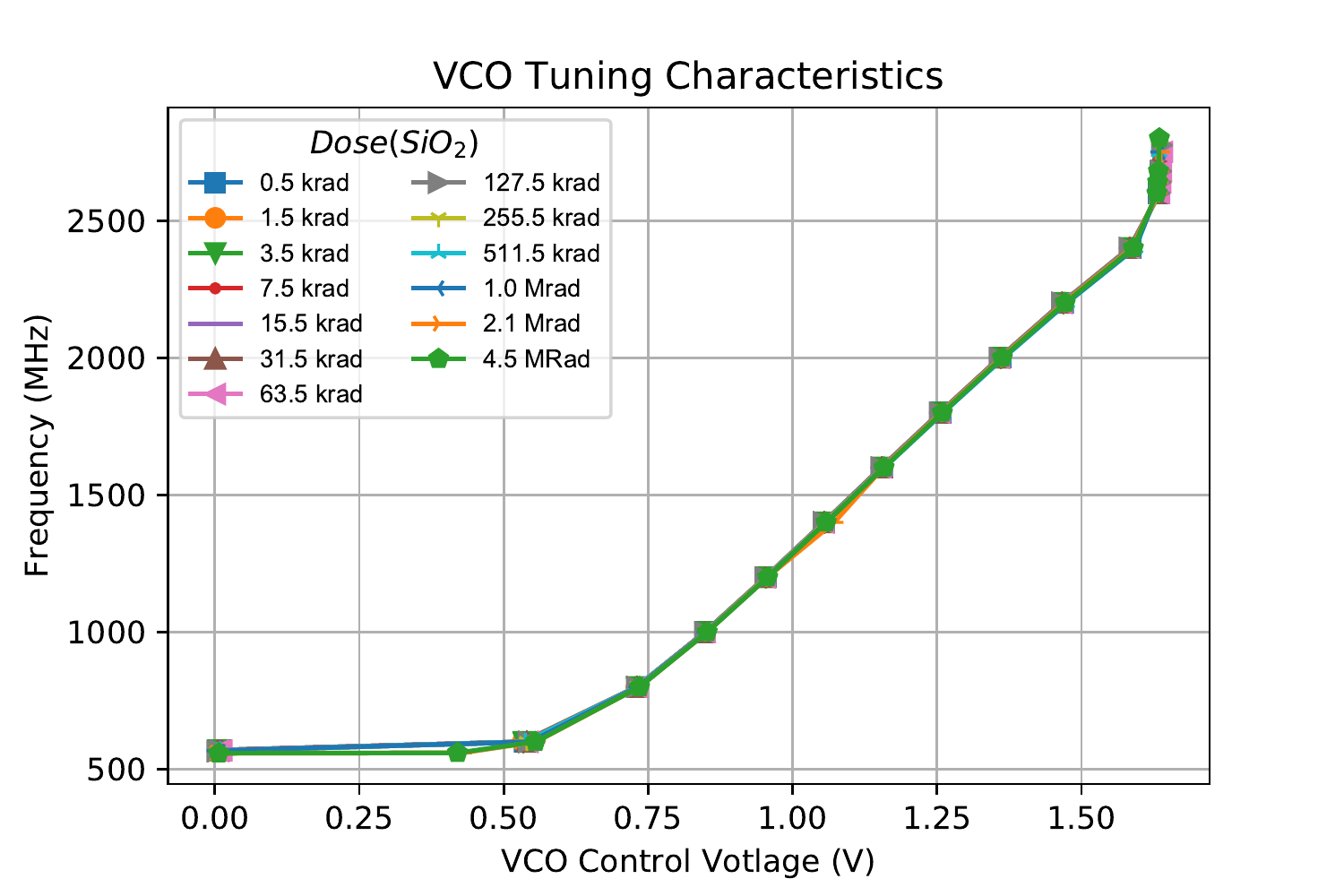}
	\caption{VCO tuning characteristics under 13 steps TID.}
	\DeclareGraphicsExtensions{.pdf,.jpeg,.png}
	\label{radvco}
\end{figure}

Irradiation test was carried out with a cabinet X-ray irradiator, X-Ray iR-160, at fixed operation potential and current. 
A transmitter chip was irradiated in 13 steps to a total dose of 4.5 Mrad ($SiO_2$). The chip was measured after each irradiation step. No annealing was performed for the first 8 steps, where the total dose is less than 128 Krad. Each of other 5 steps included 17 hours annealing. A loopback test was performed after irradiation and no transmission errors were observed. Fig. \ref{radcurrent} shows measured current on the power supplies when loopback tests were running. Analog power supplies the PLL, the CML driver and an analog buffer for the test, and the digital power supplies the serializer and all the digital circuits. There is no obvious change in the analog current. However, a peak of digital current can be seen around 500 Krad to 1 Mrad. No jitter degeneration was observed during the test. Figure \ref{radvco} shows VCO tuning characteristics on each step.

\begin{figure}[!t]
	\centering
	\includegraphics[width=1\linewidth]{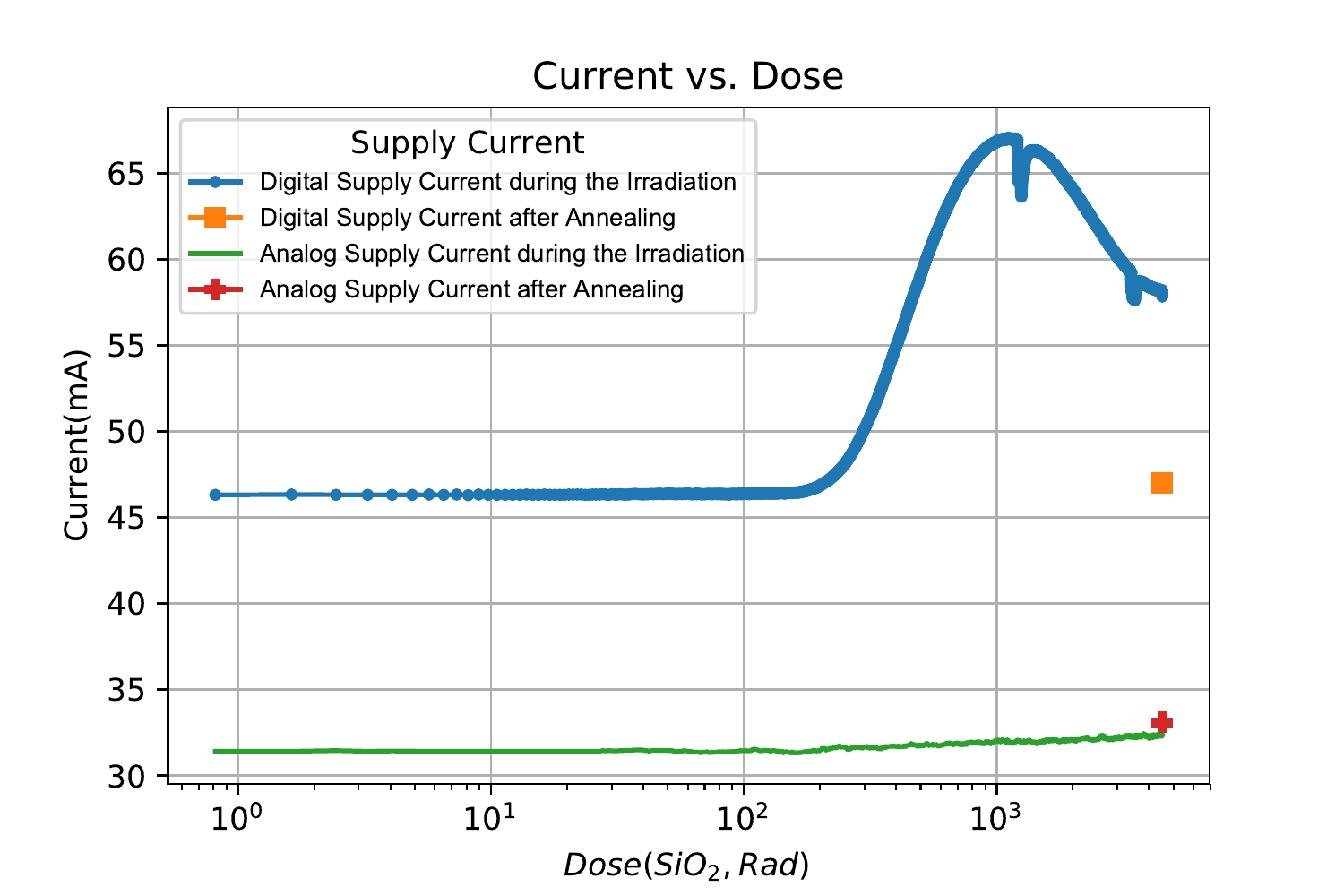}
	\caption{Real time and after annealing current consumption the transmitter in the second TID test.}
	\DeclareGraphicsExtensions{.pdf,.jpeg,.png}
	\label{rtc}
\end{figure}

In the second irradiation test, another transmitter chip was continuously irradiated to a total dose of 4.5 Mrad ($SiO_2$) while the real time current consumption of analog and digital supply are monitored. A 22-hour annealing was carried after the irradiation and the current consumptions on both supplies were measured again. The results are shown in Fig. \ref{rtc}. The device under test function normally after irradiation test.

\begin{table}[!t]
	\centering
	\caption{measured jitter of the serial data}\label{tab_jitter}
	\begin{tabular}{  c | c }
		\hline
		Jitter Classfication& Value\\ \hline 
		Random(RMS)& 3.1 ps\\
		Deterministic(PK-PK)& 25.7 ps\\
		Periodic& 16.4 ps\\
		Duty Cycle& 4.6 ps \\
		Total@10-12 BER(PK-PK)& 60.2 ps\\ \hline
	\end{tabular}
\end{table}

\section{Conclusion}
A serial link transmitter is developed aiming to meet the requirement of serial data transmission for MAPS in future subatomic physics experiments. Measurement shows the transmitter sending data at 3.2 Gb/s reliably. Reed-Solomon code provides additional robustness for data transmission. The pre-emphasis in the transmitter allows transmission over low-mass cables. The transmitter is able to withstand TID of up to at least 4.5 MRad.


%

%
%

\section*{Acknowledgment}

The authors would like to thank IPHC for supporting the prototype fabrication.

\ifCLASSOPTIONcaptionsoff
  \newpage
\fi



%
\bibliographystyle{IEEEtran}
\bibliography{mybibfile}

%
%

%
%
%
%
%




\end{document}